\documentclass[10pt, conference]{IEEEtran}
\IEEEoverridecommandlockouts
\usepackage{cite}
\usepackage{amsmath,amssymb,amsfonts}
\usepackage{graphicx}
\usepackage{xcolor}
\usepackage{hyperref}
\usepackage{tikz}
\usepackage{multirow}
\usepackage{booktabs}
\usepackage{placeins}

\usetikzlibrary{positioning}

\begin{document}

\title{When T-Depth Misleads: Predicting Fault-Tolerant Quantum Execution Slowdown under Magic-State Delivery Constraints}

\author{
Boshuai Ye\textsuperscript{1}, Arif Ali Khan\textsuperscript{1}, Peng Liang\textsuperscript{2} \\
\textsuperscript{1}\textit{M3S Research Group, University of Oulu, Oulu, Finland} \\
\textsuperscript{2}\textit{School of Computer Science, Wuhan University, Wuhan, China} \\
\{boshuai.ye, arif.khan\}@oulu.fi, liangp@whu.edu.cn
}
\maketitle

\begin{abstract}
The efficient execution of fault-tolerant quantum algorithms is fundamentally limited by the production rate of magic states required for non-Clifford operations. While circuit optimization typically targets T-depth, static T-depth does not reliably predict executable performance under bounded T-state delivery. We introduce a model that captures demand--supply imbalance using two key quantities: slack ratio, a structural indicator of scheduling flexibility, and $\Delta_{\max}$, a measure of cumulative demand surplus. We show that $\Delta_{\max}$ is a strong schedule-level indicator of execution slowdown and, together with the buffer capacity, yields a provable lower bound on executable makespan for a fixed schedule. Empirical evaluation across compressibility-controlled families and circuit workload families shows that slack ratio provides a modest but statistically supported structural signal beyond T-depth in the compressibility-family evaluation, while $\Delta_{\max}$ is the strongest schedule-level indicator of slowdown. The circuit workloads cover serial and parallel arithmetic, modular arithmetic, QAOA MaxCut, and exact quantum Fourier transform (QFT) traces. Across 4{,}904 finite schedule instances in the compressibility-family evaluation, the lower bound shows zero observed violations, with 88.9\% of instances falling within one cycle of the bound in aggregate. Notably, approximate QFT decomposition can reduce delivery pressure without altering ideal dependency depth. Compilers targeting fault-tolerant execution should treat delivery capacity as an explicit scheduling constraint.
\end{abstract}

\begin{IEEEkeywords}
fault-tolerant quantum computing, magic states, scheduling, execution modeling, performance prediction
\end{IEEEkeywords}

\section{Introduction}
\label{sec:intro}

Fault-tolerant quantum computing (FTQC) requires the use of distilled \textbf{magic states} to implement \textbf{non-Clifford operations} (such as T gates) \cite{bravyi2005universal}. The generation and delivery of these states introduce hardware-level resource constraints that are not captured by traditional, resource-agnostic circuit metrics such as T-depth \cite{amy2014polynomial, selinger2013quantum}. While T-depth measures the minimum number of sequential T-gate layers assuming infinite resources, it fails to account for the finite production rate of the magic-state factories \cite{o2017quantum, beverland2022assessing}.

In practice, circuits with lower T-depth may exhibit worse execution performance when the supply of magic states is capacity-limited. This mismatch arises from the temporal distribution of T-gate demand: a schedule that minimizes static depth can concentrate T-gate demand into \textbf{high-concurrency bursts} that exceed the available delivery rate and force execution to stall \cite{paler2019clifford}. Two schedules may have the same T-count but different temporal T-gate distributions: a depth-minimized schedule may concentrate many T gates in one layer, whereas a slightly deeper schedule may spread them more evenly and execute without delay. In an FTQC setting, such stalls translate into additional protected idle time, during which logical qubits must remain under active error correction while waiting for future magic-state availability. Prior work on fault-tolerant resource estimation and surface-code execution highlights that prolonged execution increases space--time volume and may affect reliability and resource costs~\cite{fowler2012surface,litinski2019game}. Although we do not explicitly model logical failure probabilities, delivery-induced protected idling remains undesirable from both performance and reliability perspectives.

This paper studies how bounded T-state delivery alters executable performance relative to static circuit metrics. We argue that execution degradation is driven by demand--supply imbalance, not static depth alone. We identify two quantities that characterize this imbalance at complementary levels: (i)~slack ratio, which captures structural scheduling flexibility at the circuit level, and (ii)~$\Delta_{\max}$, which measures the cumulative demand surplus under a given schedule.

We use \textbf{T-depth inversion} to refer to the schedule-comparison phenomenon in which, for two schedules $\sigma_1$ and $\sigma_2$ of the same circuit under fixed $(C,B)$, $T_{\text{static}}(\sigma_1) < T_{\text{static}}(\sigma_2)$ but $T_{\text{exe}}(\sigma_1) > T_{\text{exe}}(\sigma_2)$. In the evaluation, we instantiate this comparison using a depth-oriented reference schedule and a slack-based reshaping schedule. More generally, cumulative demand--supply imbalance is related to classical resource-constrained scheduling, but here we focus on the distinction between structural flexibility of the circuit and the delivery pressure induced by a specific execution order.

The \textbf{contributions} of this study are the following:
\begin{itemize}
\item We demonstrate that \textbf{T-depth-only} scheduling objectives can misrepresent actual executable performance under bounded T-state delivery. In the compressibility-family evaluation, slack ratio provides a modest but statistically supported structural signal beyond T-depth for predicting both execution stalls and \textbf{T-depth inversion risk} (where the shallower schedule executes more slowly).
\item We show that $\Delta_{\max}$ is the strongest schedule-level indicator of \textbf{execution stalls and makespan slowdown} in the compressibility-family evaluation.
\item We establish a provable lower bound on executable makespan and empirically assess its tightness across \textbf{4{,}904 finite schedule instances} in the compressibility-family evaluation, with zero observed violations. To support reproducibility, we release the evaluation framework, generated datasets, and analysis scripts at \url{https://github.com/C2-Q/BARC/}.
\end{itemize}

\section{Background and Related Work}
\label{sec:background}

\subsection{T Gates in Fault-Tolerant Quantum Computing}

In fault-tolerant quantum computing, the Clifford$+T$ gate set provides a universal basis for computation under quantum error correction. Clifford gates can often be implemented at relatively low logical cost, whereas the T gate requires the preparation and consumption of a distilled magic state~\cite{bravyi2005universal}. A dedicated magic-state factory must therefore continuously produce high-fidelity ancillary states, which are then consumed one per T gate through gate teleportation~\cite{gottesman1999demonstrating}. As a result, T gates are substantially more expensive than Clifford gates in both time and space, and they often dominate the cost of large fault-tolerant computations. Magic-state distillation protocols have been further optimized to reduce overhead~\cite{bravyi2012magic}.

\subsection{T-Count and T-Depth as Compilation Objectives}

Because T gates dominate fault-tolerant cost, many compilation methods aim to reduce either the total number of T gates or the number of sequential T-gate layers~\cite{amy2014polynomial,amy2019t,vandaele2025lower}. The latter quantity is commonly referred to as T-depth. Recent LUT-based synthesis work for Boolean circuits further reduces T-count using relative-phase implementations and Boolean decompositions~\cite{clarino2024leveraging}. Such techniques reduce non-Clifford demand, but they do not indicate whether the remaining T gates are scheduled in bursts that exceed a bounded delivery rate. T-depth minimization is justified when T-states can be supplied on demand: under that assumption, a circuit with smaller T-depth should execute faster.

In practice, however, T-state availability may be limited by production and delivery constraints. T-depth describes the logical layering of T gates, but it does not describe whether the corresponding T-states can be produced and delivered at the rate implied by the compiled execution order. Once T-state delivery becomes capacity-limited, static depth and executable runtime need not coincide.

\subsection{Magic-State Delivery as an Architectural Constraint}

Fault-tolerant resource-estimation and architecture studies treat magic-state production as a constrained resource. Factory throughput, factory count, and qubit overhead are common parameters in design-space exploration and cost estimation~\cite{o2017quantum,van2023using,beverland2022assessing}. Multi-level factory optimization also frames magic-state production as a supply-chain problem between factory zones and the logical processor, exposing space--time tradeoffs between factory investment and execution time~\cite{silva2024optimizing}. Surface-code studies also show that runtime depends on architectural space--time tradeoffs, not logical depth alone~\cite{litinski2019game}. Time-efficient fault-tolerant computation links reduced execution time to additional space overhead~\cite{yamasaki2024time}.

From a software perspective, delivery-constrained T-gate execution can be viewed as a specialized precedence- and resource-constrained scheduling problem, closely related to the Resource-Constrained Project Scheduling Problem (RCPSP)~\cite{hartman2011survey}. Modern compilation frameworks such as Qiskit~\cite{javadi2024quantum} and tket~\cite{sivarajah2021t} include sophisticated heuristics for qubit mapping, routing, and hardware-aware compilation, with objectives shaped primarily by device connectivity, routing overhead, and circuit depth in near-term hardware settings. These compilation flows do not explicitly model bounded magic-state delivery as a scheduling constraint for fault-tolerant execution.

A related line of work has considered scheduling and queueing effects under restricted hardware resources. Paler and Basmadjian~\cite{paler2019clifford}, for instance, study T-gate scheduling using queueing models for topological assemblies and observe that some circuits, such as adders, may not experience increased execution time even under hardware restrictions. Related work has also addressed the mapping and scheduling of magic-state distillation circuits themselves, including optimizations for multi-level factory structures in fault-tolerant architectures~\cite{ding2018magic}. These studies show that circuit structure and constrained supply affect execution behavior, but emphasize architecture provisioning, factory scheduling, or schedule construction.

For compiler analysis, a useful intermediate step is a lightweight diagnostic for fixed or compiler-produced schedules: which circuits have enough structural freedom to reshape T demand, and which schedules create delivery pressure under bounded T-state supply? We study this through two quantities: slack ratio for structural flexibility and $\Delta_{\max}$ for schedule-level demand--supply imbalance.

\section{Execution Model}
\label{sec:model}

We study the execution of a compiled quantum circuit under bounded T-state delivery, where the number of T-states that can be supplied over time is limited. Our goal is to describe how this constraint affects the executable makespan of a fixed schedule.

\subsection{Demand, Supply, and Stall}

We represent the circuit as a directed acyclic graph (DAG), in which nodes are operations and edges represent dependency constraints. A schedule $\sigma$ assigns operations to logical time steps while respecting these dependencies. Given a fixed schedule $\sigma$, we define the T-demand trace $D_\sigma(t)$ as the number of T gates scheduled at time step $t$. The cumulative demand up to time step $t$ is
\begin{equation}
A_\sigma(t) = \sum_{\tau \le t} D_\sigma(\tau).
\end{equation}
We index logical time steps as $t=1,\ldots,T_{\text{static}}$, and assume that the $C$ T-states produced during a logical step are available for the T gates executed at that step. We model T-state availability using two parameters: a delivery capacity $C$ and a buffer capacity $B$. Here, $C$ denotes the maximum number of T-states produced per time step, while $B$ denotes the maximum number of T-states stored in advance. We assume that the buffer is initially full. Thus, at the beginning of execution the system may hold up to $B$ pre-produced T states, and during each logical step at most $C$ additional T states become available. Here $D_\sigma(t) \in \mathbb{Z}_{\ge 0}$ and $B \in \mathbb{Z}_{\ge 0}$ are measured in logical T-state units, while $C \in \mathbb{Z}_{>0}$ is measured in logical T states per logical time step. Under this model, the cumulative supply available up to time step $t$ is bounded by
\begin{equation}
S(t) = Ct + B.
\end{equation}

If the cumulative demand exceeds the cumulative supply, the scheduled T-gate demand at that logical step cannot be executed when planned because the required T-states are not yet available. The execution must then wait for future T-state production. We refer to this waiting as a \emph{stall}, and to the unmet cumulative demand as \emph{backlog}. Any contiguous interval of logical steps for which $A_\sigma(t) > Ct + B$ holds is a \emph{backlog-active interval}.

\begin{figure}[t]
  \centering
  \includegraphics[width=0.95\columnwidth]{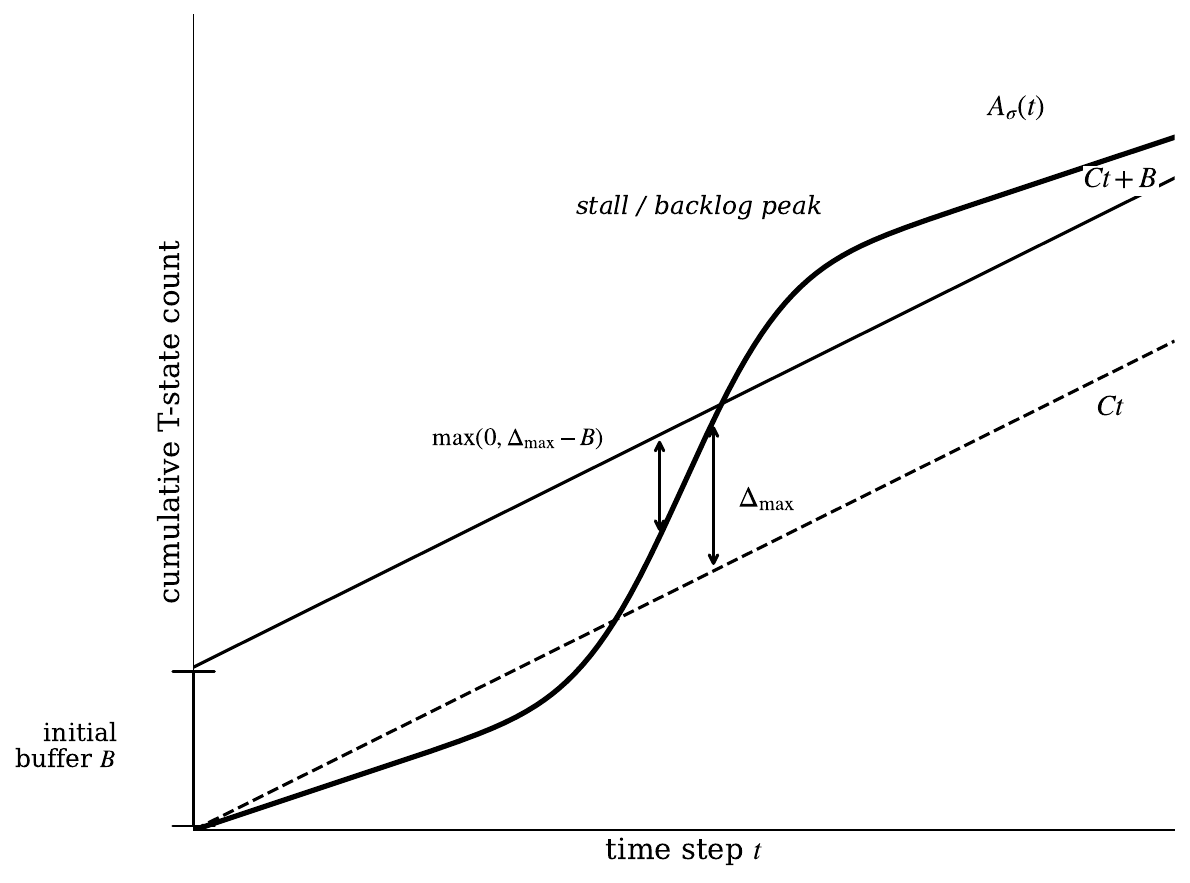}
  \caption{Cumulative T-state demand $A_{\sigma}(t)$, production line $Ct$, and bounded available supply $Ct+B$. Here $\Delta_{\max}$ denotes the maximum gap between $A_{\sigma}(t)$ and $Ct$, whereas $P_B(\sigma)=\max(0,\Delta_{\max}-B)$ denotes the buffer-adjusted surplus above $Ct+B$.}
  \label{fig:delta_max_illustration}
\end{figure}

Figure~\ref{fig:delta_max_illustration} illustrates this model. The curve $A_\sigma(t)$ shows cumulative demand, while the line $Ct+B$ shows the maximum cumulative supply, where the intercept $B$ corresponds to the initially available buffer. Whenever the demand curve rises above the supply curve, backlog accumulates and execution is delayed. In the present trace-level execution model, each logical time step is treated as an atomic demand unit: the T gates assigned to a given step must be executable together when that step is reached, without splitting that step's demand across later logical steps. Under this semantics, a logical step whose demand exceeds the maximum simultaneously available stock, $B+C$, cannot be executed as scheduled. Equivalently, if $D_\sigma(t) > B + C$ for some $t$, we mark that schedule instance infeasible under the trace-level model and set $T_{\text{exe}}=\infty$. Such cases correspond to schedules that would require finer-grained splitting or rescheduling beyond the fixed-step model. The finite-instance analysis in the remainder of the paper is therefore conditional on this atomic-step delivery semantics.

\subsection{Demand--Supply Imbalance and Lower Bound}
\label{sec:model_delta}

To quantify demand--supply mismatch, we define
\begin{equation}
\Delta_{\max}(\sigma) = \max_t \left( A_\sigma(t) - Ct \right).
\end{equation}

This quantity measures the largest cumulative surplus of demand over production capacity. Equivalently, it is the maximum vertical gap between $A_\sigma(t)$ and the production line $Ct$ in Fig.~\ref{fig:delta_max_illustration}. We therefore refer to $\Delta_{\max}$ as the unbuffered cumulative surplus. For a fixed buffer capacity $B$, the buffer-adjusted surplus is
\begin{equation}
P_B(\sigma)=\max(0,\Delta_{\max}(\sigma)-B).
\end{equation}
This quantity measures the cumulative surplus that remains after the buffer has been accounted for.

Let $T_{\text{static}}$ denote the number of logical time steps in the given schedule when T-state delivery is assumed to be unconstrained, and let $T_{\text{exe}}$ denote the actual execution time under bounded T-state delivery. The following lower bound holds for any fixed schedule:
\begin{equation}
T_{\text{exe}} \ge T_{\text{static}} +
  \left\lceil \frac{P_B(\sigma)}{C} \right\rceil.
\label{eq:bound}
\end{equation}

The bound follows from the fact that cumulative supply up to time step $t$ is bounded by $S(t)=Ct+B$, which includes the contribution of the initially available buffer. When $P_B(\sigma)>0$, the remaining surplus must be cleared by future T-state production. Since T-states are produced at rate $C$, clearing this surplus requires at least $\lceil P_B(\sigma)/C \rceil$ additional logical cycles. In an FTQC setting, these extra cycles correspond to additional protected logical cycles during which the system must continue error-corrected execution while waiting for future T-state availability. This bound is schedule-specific and applies to the supplied execution order.

\subsection{Structural Flexibility}
\label{sec:model_slack}

Not all circuits impose the same scheduling rigidity. To describe how much freedom a circuit offers for relocating T gates without violating dependencies, we use slack. For each T gate $v$, we define
\begin{equation}
\mathrm{slack}(v) = LS(v) - ES(v),
\end{equation}
where $ES(v)$ is the earliest start time of $v$ under an ASAP (as-soon-as-possible) traversal of the dependency DAG, and $LS(v)$ is the latest start time of $v$ under the corresponding ALAP (as-late-as-possible) traversal that preserves the minimum unconstrained static schedule length. Thus, $\mathrm{slack}(v)$ is the standard ASAP/ALAP slack relative to the critical-path makespan of the DAG, following conventional precedence-constrained scheduling terminology.

To summarize this at the circuit level, we use the \emph{slack ratio}: the fraction of T gates with positive slack. This is a \emph{structural} slack measure derived from precedence flexibility in the DAG. It summarizes how much freedom the circuit provides for reshaping T-gate demand while remaining relatively stable across circuits of different sizes.

\subsection{Model Assumptions, Scope, and Complexity}
We assume deterministic delivery and capacity-limited buffering. The model does not include stochastic factory failures, storage decay, routing overhead, or control latency, and it omits spatial effects such as placement and communication distance. The model therefore isolates temporal demand--supply mismatch as the primary driver of makespan inflation.

This level of abstraction is consistent with prior fault-tolerant resource-estimation work~\cite{van2023using,beverland2022assessing} and with fault-tolerant architecture studies that model magic-state production capacity at the logical level~\cite{o2017quantum,silva2024optimizing}.

Both proposed metrics are computationally inexpensive and suitable for integration into existing compiler passes. Computing $\Delta_{\max}$ requires a single pass over the T-demand trace of a candidate schedule, yielding $O(N_T)$ complexity where $N_T$ is the T-count of the circuit. Computing the slack ratio requires standard critical-path analysis on the circuit DAG, which runs in $O(|V|+|E|)$ time where $|V|$ and $|E|$ are the number of nodes and edges, respectively. Neither metric requires re-compilation or simulation of the full quantum execution. 

\section{Experimental Setup}
\label{sec:setup}
We evaluate the proposed execution model and predictor framework using two complementary workload groups. The compressibility-controlled families provide a controlled setting for predictor analysis and lower-bound validation, with systematically varied scheduling flexibility. The circuit workload families examine delivery pressure in concrete arithmetic, modular-arithmetic, optimization, and transform workloads. We compare three scheduling policies with distinct roles: a depth-oriented reference, a simple resource-aware baseline, and a slack-based reshaping intervention. Tables~\ref{tab:workloads} and~\ref{tab:policies} summarize the evaluated workloads and policies.

\begin{table}[t]
\caption{Workload families used in the evaluation.}
\label{tab:workloads}
\centering
\scriptsize
\setlength{\tabcolsep}{2.2pt}
\begin{tabular}{p{1.25cm} p{1.75cm} p{2.05cm} p{2.35cm}}
\hline
Class & Family & Size / Range & Role \\
\hline
Compress. & High & 5 seeds; $C=1$--$7$; $B=0$--$15$ & High scheduling flexibility; exposes inversion and reshaping effects \\
Compress. & Medium & 5 seeds; $C=1$--$7$; $B=0$--$15$ & Intermediate scheduling flexibility; transition regime \\
Compress. & Low & 5 seeds; $C=1$--$7$; $B=0$--$15$ & Low scheduling flexibility; control regime \\
Arithmetic & Ripple adder & $n \in \{4,6,8,12,16\}$ & Serial arithmetic baseline \\
Arithmetic & CLA adder & $n \in \{4,8,12,16\}$ & Parallel-prefix addition workload \\
Arithmetic & Multiplier & $n \in \{4,5,6,7,8,12,16\}$ & Arithmetic workload with increasing delivery pressure \\
Mod. arith. & Mod. arith. block & $n \in \{4,6,8\}$ & Building-block workload for modular-arithmetic patterns \\
Optimization & QAOA MaxCut & $n \in \{6,8,10\}$; $p \in \{1,2\}$ & Graph-structured optimization workload \\
Transform & Exact QFT & $n \in \{8,12,16\}$ & High-pressure transform workload; full scan at $n=8$ \\
Approx. study & QFT ($n=12$) & exact and degree-4 approx.; reduced grid & Effect of approximation on delivery pressure \\
\hline
\end{tabular}
\end{table}

\begin{table}[t]
\caption{Scheduling policies used in the evaluation.}
\label{tab:policies}
\centering
\begin{tabular}{p{2.0cm} p{5.1cm}}
\hline
Policy & Role in analysis \\
\hline
$\sigma_{\text{static}}$ & Depth-oriented reference policy that greedily schedules ready T gates as early as possible. \\
$\sigma_{\text{ca}}$ & Resource-aware baseline that enforces a per-layer quota equal to delivery capacity $C$, without buffer-aware lookahead. \\
$\sigma_{\text{smooth}}$ & Slack-based reshaping heuristic that spreads T-gate demand more evenly while respecting dependencies. \\
\hline
\end{tabular}
\end{table}

\subsection{Workloads}
\label{sec:setup_workloads}

To vary structural flexibility in a controlled way, we construct three families of dependency-aware DAGs, following the common use of synthetic precedence-constrained instances to isolate scheduling effects~\cite{hartman2011survey}. These families differ in the density of precedence constraints among T gates. High-compressibility instances use sparse T-gate dependencies and therefore admit substantial slack, low-compressibility instances use dense T-gate dependencies and therefore admit little slack, and medium-compressibility instances fall between these extremes. Each instance is generated from a layered dependency template: we instantiate a layered DAG skeleton and sample precedence edges between pairs of T gates subject to acyclicity and layer order, with sampling density determined by the target family. The five reported seeds correspond to independent random instances generated within each family.

Each compressibility family is evaluated across five random seeds and a grid of delivery parameters $C \in \{1,\ldots,7\}$ and $B \in \{0,\ldots,15\}$. Across the three families, this yields 1{,}680 workload settings; evaluated under the three scheduling policies in Table~\ref{tab:policies}, the compressibility-family evaluation contains 5{,}040 schedule instances. Of these, 4{,}904 are finite executable instances used in lower-bound validation; the remaining 136 are infeasible under the atomic-step delivery model and are recorded as $T_{\text{exe}}=\infty$.

The second group consists of circuit workload families derived from Clifford$+T$ traces: ripple-carry addition, carry-lookahead addition, multiplication, a modular-arithmetic block, QAOA MaxCut, and exact QFT. These families cover common arithmetic, optimization, and transform structures used in quantum algorithm design, including reversible arithmetic circuits~\cite{vedral1996quantum,cuccaro2004new,draper406142logarithmic}, QAOA-style MaxCut circuits~\cite{farhi2014quantum}, and QFT-based transform workloads~\cite{coppersmith2002approximate,nielsen2010quantum}. These workloads show where concrete algorithmic structures fall in the structural--system space defined by slack ratio and $\Delta_{\max}$. The modular-arithmetic block is used as a building-block workload representative of modular-arithmetic patterns. QAOA is evaluated on a reduced $(C,B)$ grid because the chosen arbitrary-rotation synthesis configuration produces long Clifford$+T$ traces. The reduced QAOA grid uses $C \in \{1,2,3,5,7\}$ and $B \in \{0,4,8,12,15\}$. At each logical time step, we count T and T$^\dagger$ gates and use these counts as the T-demand trace for bounded-delivery simulation. For larger exact-QFT instances, Clifford$+T$ synthesis produces very long T-demand traces, making exhaustive evaluation over the full $(C,B)$ grid computationally expensive. We therefore include those instances only in the structural analysis and reserve full bounded-delivery scans for the largest tractable exact-QFT instance.

We also evaluate one algorithmic transformation study on QFT at $n=12$, comparing exact and approximate decompositions on a reduced $(C,B)$ grid to isolate whether approximation changes ideal circuit structure or reduces bounded-delivery pressure at fixed dependency depth.

The simulation framework, DAG generators, and evaluated instance tables are available at \url{https://github.com/C2-Q/BARC/}.

\subsection{Scheduling Policies and Their Roles}
\label{sec:setup_policies}

The static-min policy $\sigma_{\text{static}}$ represents a compiler that minimizes static T-depth without modeling delivery constraints. The capacity-aware policy $\sigma_{\text{ca}}$ accounts for delivery capacity through a per-layer quota equal to $C$, but intentionally does not exploit dependency-derived slack or buffer-aware lookahead. The smoothed policy $\sigma_{\text{smooth}}$ is a slack-based reshaping heuristic: it computes ASAP/ALAP slack on the circuit DAG, introduces a small slack-derived extension to the minimum unconstrained makespan, and then schedules ready T gates using relaxed deadlines and a roughly even per-cycle T-demand target while respecting dependencies. Its purpose is not to establish a universally dominant heuristic, but to expose the role of structural flexibility. 

\subsection{Evaluation Metrics and Protocol}
We report the following quantities throughout the evaluation.

\paragraph{Static schedule length ($T_{\text{static}}$)}
The number of logical time steps in the schedule when T-state delivery is assumed to be unconstrained.

\paragraph{Executable makespan ($T_{\text{exe}}$)}
The actual number of logical time steps required to complete execution under bounded T-state delivery, including any stall cycles introduced by supply shortfalls. Infeasible instances under the atomic-step delivery model are recorded as $T_{\text{exe}}=\infty$.

\paragraph{Slowdown ratio}
The ratio $T_{\text{exe}} / T_{\text{static}}$, measuring execution inflation relative to the static schedule.

\paragraph{Stall}
We record both a binary indicator of whether any stall cycle occurs during execution and the total number of stall cycles incurred.

\paragraph{Slack ratio}
The fraction of T gates with positive slack, as defined in Section~\ref{sec:model_slack}. This is our pre-scheduling structural indicator of scheduling flexibility under bounded-delivery execution.

\paragraph{$\Delta_{\max}$}
The unbuffered cumulative surplus under the given schedule, as defined in Section~\ref{sec:model_delta}. For a fixed buffer capacity $B$, we use $P_B(\sigma)=\max(0,\Delta_{\max}(\sigma)-B)$ as the buffer-adjusted surplus entering the execution lower bound.

\paragraph{Lower-bound prediction} The fixed-schedule executable-makespan lower bound from Eq.~\eqref{eq:bound}, used to assess how tightly $P_B(\sigma)$ constrains the actual executable makespan.

We evaluate the diagnostic value of these indicators primarily on the compressibility-family evaluation using three tasks:
\begin{itemize}
\item \textbf{Stall classification:} Given a circuit and a $(C,B)$ pair, determine whether execution incurs any stall. We evaluate this task using the area under the receiver operating characteristic curve (AUC).
\item \textbf{Slowdown regression:} Estimate the slowdown ratio $T_{\text{exe}} / T_{\text{static}}$ from structural and schedule-level quantities. We evaluate this task primarily using Spearman rank correlation, denoted by $\rho$, because the relationships of interest are monotonic but not necessarily linear; when linear agreement is specifically relevant, we report Pearson correlation, denoted by $r$.
\item \textbf{T-depth inversion classification:} Determine whether $\sigma_{\text{static}}$ yields a longer executable makespan than $\sigma_{\text{smooth}}$ despite having a shorter static schedule. We evaluate this task using AUC.
\end{itemize}

For the comparison between slack ratio and T-depth, we additionally report paired bootstrap confidence intervals (10{,}000 resamples) for the difference in predictive performance. For each finite schedule instance in the compressibility-family evaluation, we also compute the predicted lower bound from Eq.~\eqref{eq:bound} and compare it with the observed $T_{\text{exe}}$. We report the number of bound violations, the Pearson correlation between predicted and observed makespans, and the distribution of the gap $T_{\text{exe}} - \text{lower bound}$, along with a post-hoc analysis of which circuit and system properties are associated with larger gaps.

\section{Results and Analysis}

\subsection{Static Depth Does Not Predict Executable Performance}
\label{sec:results_mismatch}
Shorter static schedules do not necessarily yield faster execution under bounded T-state delivery. Across the compressibility-family evaluation, circuits in the high-compressibility family exhibit substantial execution inflation under $\sigma_{\text{static}}$ despite having the shortest static schedules. The low-compressibility family, by contrast, shows negligible slowdown under identical delivery conditions. This divergence reflects demand concentration: static depth measures how many T layers are present, but not whether those layers exceed the delivery envelope.

T-depth inversion is directly visible in the high-compressibility family: 35.7\% of all evaluated $(C,B)$ configurations produce instances in which $\sigma_{\text{static}}$ has a strictly shorter static schedule than $\sigma_{\text{smooth}}$, yet a longer executable makespan. Among the subset where both schedules are feasible, this fraction rises to 41.8\%. The inversion is essentially absent in the low-compressibility family, indicating that inversion requires structural freedom: without movable T gates, bounded delivery may slow execution, but it has little room to reverse the ordering between schedules.

We also find that $\sigma_{\text{ca}}$ removes most of the direct performance gap relative to $\sigma_{\text{static}}$ in the high- and medium-compressibility families, with mean slowdown ratio approaching~1.0. The failure mode is therefore not large T-count alone, but the combination of depth-oriented scheduling and bounded delivery. We group delivery capacities into low-, mid-, and high-capacity regimes, corresponding to $C \in \{1,2\}$, $C \in \{3,4,5\}$, and $C \in \{6,7\}$, respectively. Table~\ref{tab:slowdown_ratio} shows the same pattern across capacity regimes: under $\sigma_{\text{static}}$, high-compressibility circuits slow down most strongly in the supply-limited regime, medium-compressibility circuits show a weaker version of the same trend, and low-compressibility circuits remain near~1.0 throughout.

\begin{table}[htbp]
\caption{Mean slowdown ratio under $\sigma_{\text{static}}$ by compressibility family and capacity regime.}
\label{tab:slowdown_ratio}
\centering
\begin{tabular}{@{}lccc@{}}
\toprule
\textbf{Family} & \textbf{Low Cap.} & \textbf{Mid Cap.} & \textbf{High Cap.} \\
\midrule
High-compress. & 2.96$\times$ & 1.30$\times$ & 1.01$\times$ \\
Mid-compress.  & 1.44$\times$ & 1.01$\times$ & 1.00$\times$ \\
Low-compress.  & 1.00$\times$ & 1.00$\times$ & 1.00$\times$ \\
\bottomrule
\end{tabular}
\end{table}

\subsection{Structural and System-Level Predictors}
\label{sec:results_predictors}
\begin{figure}[t]
  \centering
  \includegraphics[width=1.1\columnwidth]{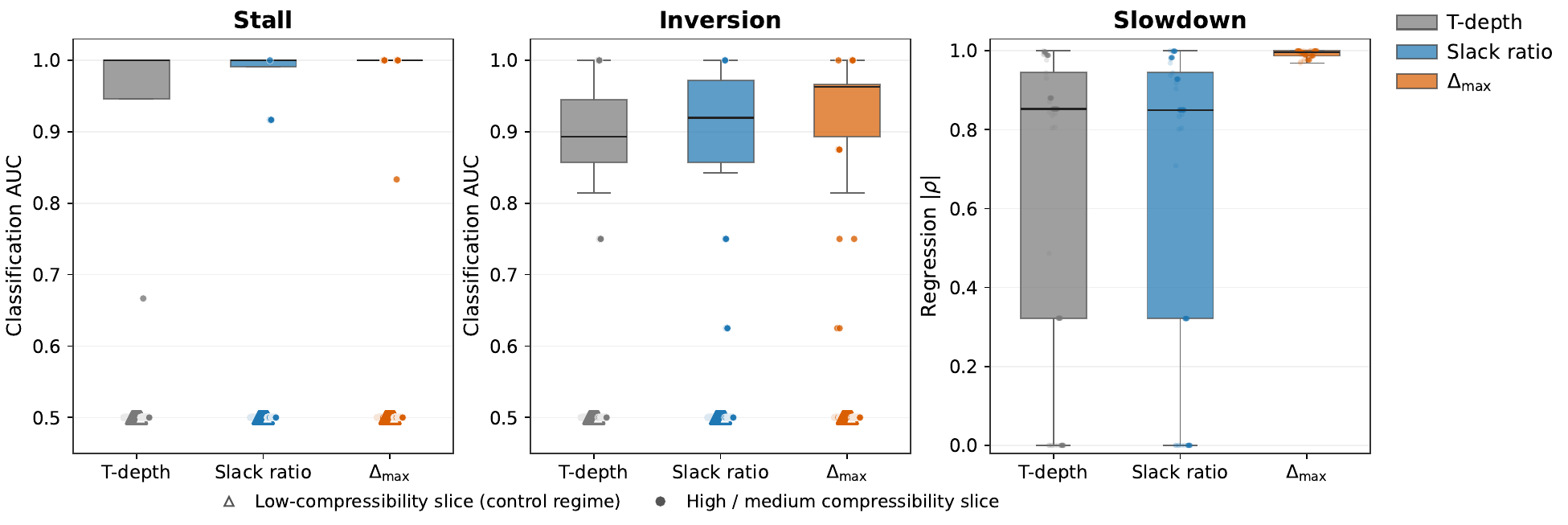}
    \caption{Predictive performance of T-depth, slack ratio, and $\Delta_{\max}$ in the compressibility-family evaluation. AUC is reported for stall and T-depth inversion classification; Spearman $|\rho|$ is reported for slowdown regression. Markers show individual $(C,B)$ slices: triangles denote low-compressibility control slices where structural variation is largely absent, and circles denote high- and medium-compressibility slices. $\Delta_{\max}$ is the strongest and most stable schedule-level predictor, while slack ratio provides a modestly stronger structural signal than T-depth in the non-degenerate regimes.}
  \label{fig:predictor_comparison}
\end{figure}

Figure~\ref{fig:predictor_comparison} reports slice-level predictive performance across three tasks: stall classification, slowdown regression, and T-depth inversion classification. Among the structural indicators in this controlled evaluation, slack ratio is modestly but consistently more informative than T-depth in the non-degenerate regimes. Using paired bootstrap resampling over evaluation slices, the average within-slice AUC advantage of slack ratio over T-depth is 0.007 for stall classification (95\%~CI [0.002, 0.012]) and 0.013 for T-depth inversion classification (95\%~CI [0.008, 0.018]). For slowdown regression the effect is in the same direction but does not reach statistical significance (95\%~CI~[$-$0.002, 0.014]).

The improvement is modest, but slack ratio reflects a structural property that T-depth does not capture: the freedom to redistribute T-gate demand across time steps. This difference is most visible in the high- and medium-compressibility regimes, where reordering freedom exists and burst shaping matters. In the low-compressibility control regime, structural variation is largely absent and all predictors approach non-informative behavior, as shown by the control-slice markers in Fig.~\ref{fig:predictor_comparison}.

\begin{figure}[t]
  \centering
  \includegraphics[width=\columnwidth]{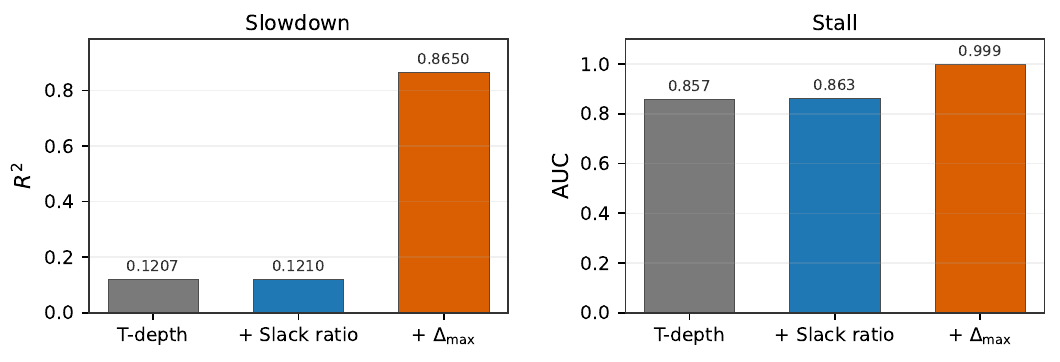}
  \caption{Incremental predictive gain from adding structural and system-level indicators in the compressibility-family evaluation. Slack ratio adds a small improvement over T-depth alone; $\Delta_{\max}$ yields a much larger increase for both slowdown and stall prediction.}
  \label{fig:incremental_gain}
\end{figure}

Figure~\ref{fig:incremental_gain} adds predictors incrementally. For slowdown prediction, a linear model using T-depth alone achieves $R^2 = 0.1207$. Adding slack ratio increases $R^2$ only slightly to 0.1210, indicating that slack contributes limited but nonzero structural information beyond T-depth alone. Adding $\Delta_{\max}$ then raises $R^2$ to 0.865, showing that most explanatory power lies at the schedule level rather than in static structure. The same pattern holds for stall prediction: AUC increases from 0.8568 with T-depth alone to 0.8627 after adding slack ratio, and then to 0.9993 after adding $\Delta_{\max}$.

\begin{figure}[t]
  \centering
  \includegraphics[width=\columnwidth]{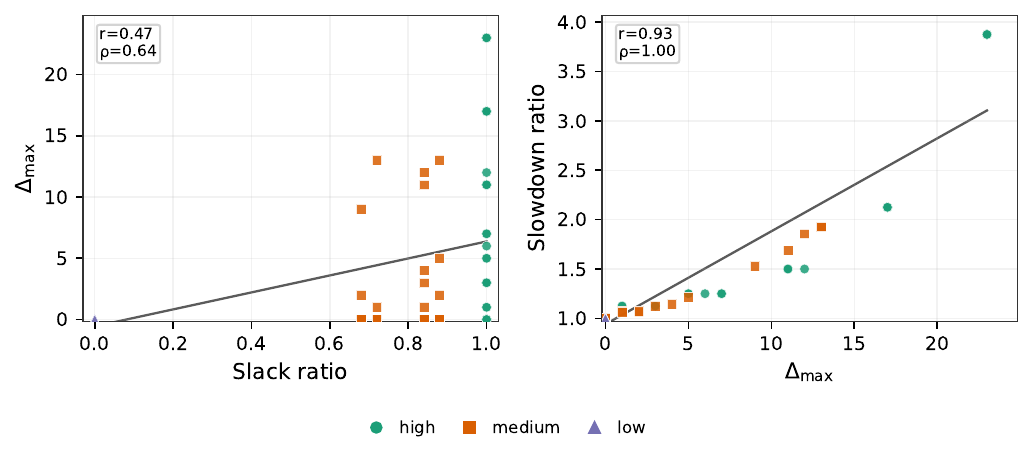}
  \caption{Empirical validation of the structure--system--execution chain in the compressibility-family evaluation under $\sigma_{\text{static}}$ at $B=6$. (Left) Slack ratio versus $\Delta_{\max}$: structural flexibility is associated with delivery pressure across capacity settings. (Right) $\Delta_{\max}$ versus slowdown ratio: delivery pressure strongly predicts execution slowdown. Panel annotations report Pearson correlation ($r$) and Spearman rank correlation ($\rho$).}

  \label{fig:empirical_chain}
\end{figure}

\begin{figure}[t]
  \centering
  \includegraphics[width=\columnwidth]{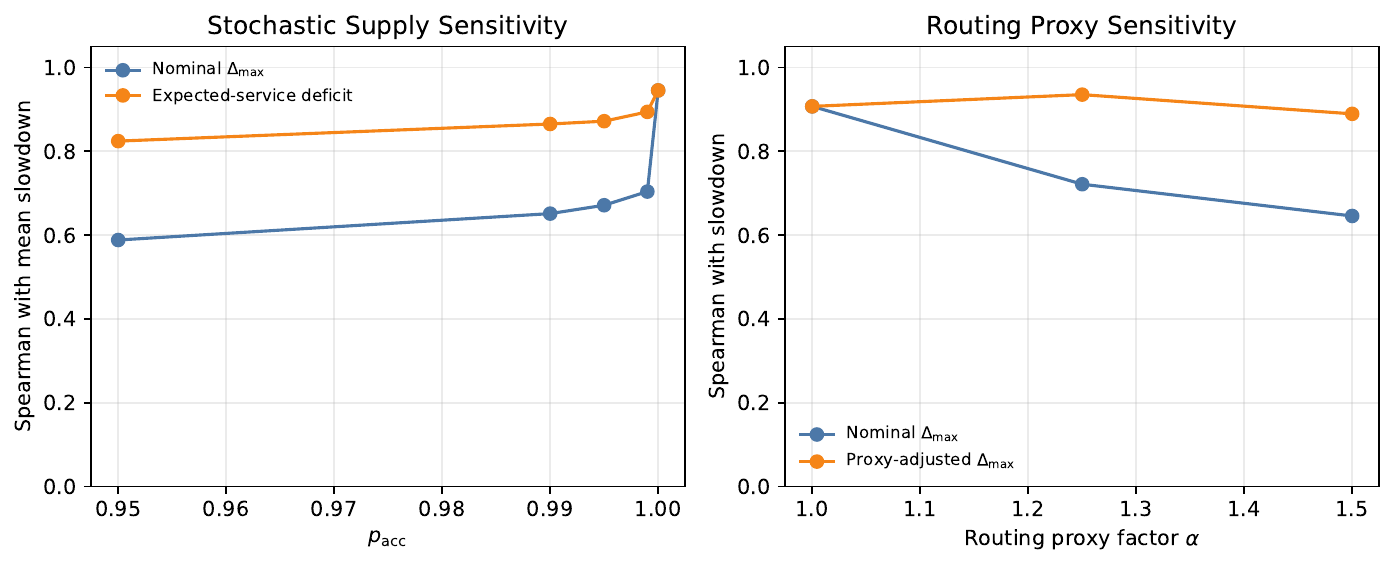}
  \caption{Sensitivity of delivery-pressure predictors beyond the deterministic main model in the compressibility-family evaluation. Left: under stochastic supply variability, the deterministic $\Delta_{\max}$ weakens as acceptance probability decreases, whereas an expected-service variant remains more stable. Right: under a route-induced effective-capacity proxy, recomputing the deficit with reduced effective capacity preserves strong association with slowdown.}
  \label{fig:appendix_robustness_supply_routing}
\end{figure}

Figure~\ref{fig:empirical_chain} shows the same ordering in a representative fixed-buffer regime ($B=6$) under $\sigma_{\text{static}}$. The left panel relates slack ratio to $\Delta_{\max}$, indicating that structural flexibility shapes delivery pressure across capacity settings. The right panel relates $\Delta_{\max}$ to slowdown ratio and remains nearly perfectly monotonic in this fixed-buffer regime. Slack ratio acts as an upstream structural signal, while $\Delta_{\max}$ serves as a model-derived schedule diagnostic for cumulative delivery pressure. Detailed subgroup stability results are reported in Appendix~\ref{app:stability}; supplementary sensitivity results for stochastic supply and route-induced capacity proxies are shown in Fig.~\ref{fig:appendix_robustness_supply_routing}.

\subsection{Empirical Tightness of the Lower Bound}
\label{sec:results_bound}
We empirically assess the tightness of the fixed-schedule lower bound from Eq.~\eqref{eq:bound} across 4{,}904 finite executable schedule instances spanning all three compressibility families, five random seeds, and a grid of $(C,B)$ parameter combinations.

\begin{figure}[t]
  \centering
  \includegraphics[width=\columnwidth]{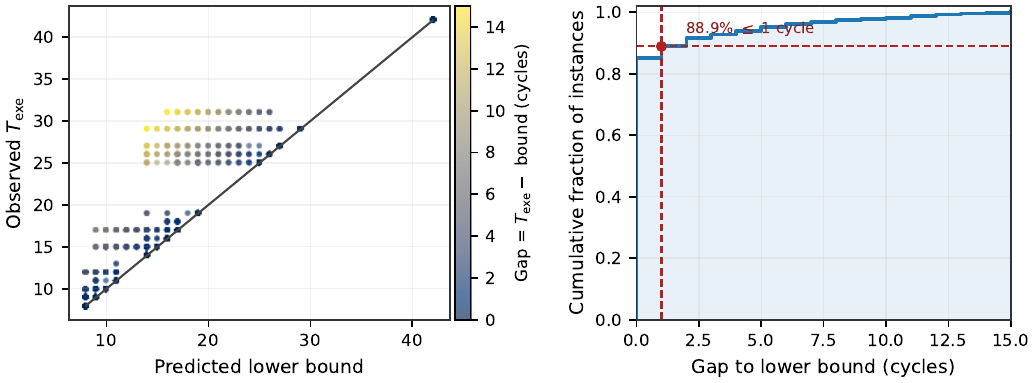}
  \caption{Predicted lower bound versus observed $T_{\text{exe}}$ across 4{,}904 finite executable schedule instances in the compressibility-family evaluation. Left: all points lie on or above the identity line, showing zero observed violations. Right: empirical cumulative distribution of the gap $T_{\text{exe}}-\text{bound}$; 88.9\% of finite instances fall within one cycle of the lower bound.}
  \label{fig:lower_bound}
\end{figure}

Zero instances violate the bound. Figure~\ref{fig:lower_bound} compares the predicted lower bound with the observed executable makespan $T_{\text{exe}}$; all points lie on or above the identity line. Over these 4{,}904 finite schedule instances, the Pearson correlation between the predicted bound and observed $T_{\text{exe}}$ is 0.988, the median gap $T_{\text{exe}}-\text{bound}$ is 0 cycles, 88.9\% of instances fall within 1 cycle of the bound, and the mean gap is 0.68 cycles.

These aggregate statistics are dominated by instances where the buffer-adjusted surplus is zero, i.e., $P_B(\sigma)=0$ or equivalently $\Delta_{\max} \leq B$. On the nontrivial subset with $P_B(\sigma)>0$ (382 instances), the bound remains valid but is less tight: the mean gap is 5.81 cycles, the median gap is 5 cycles, only 9.69\% of instances fall within 1 cycle of the bound, and the Pearson correlation is 0.857. Eq.~\eqref{eq:bound} is therefore best interpreted as a valid fixed-schedule lower bound whose tightness depends on whether backlog persists across multiple logical steps.

Within the nontrivial subset $P_B(\sigma)>0$, higher delivery capacity $C$ reduces the gap between the observed makespan and the lower bound (Mann--Whitney rank-biserial effect size $-$0.569). Circuit length $T_{\text{static}}$ also negatively predicts gap size (effect size $-$0.640), consistent with longer circuits having more opportunity to absorb transient backlogs.

Instances with larger positive gaps share a common pattern: the backlog-active interval persists across multiple logical steps instead of being concentrated near a single boundary, causing the first-order backlog term to slightly underestimate the actual delay. Representative positive-gap cases are analyzed in Appendix~\ref{app:gap_cases} and illustrated in Fig.~\ref{fig:appendix_gap_cases}.

\subsection{Circuit Workload Families and Algorithmic Implications}
\label{sec:results_real}
\begin{figure}[t]
  \centering
  \includegraphics[width=\columnwidth]{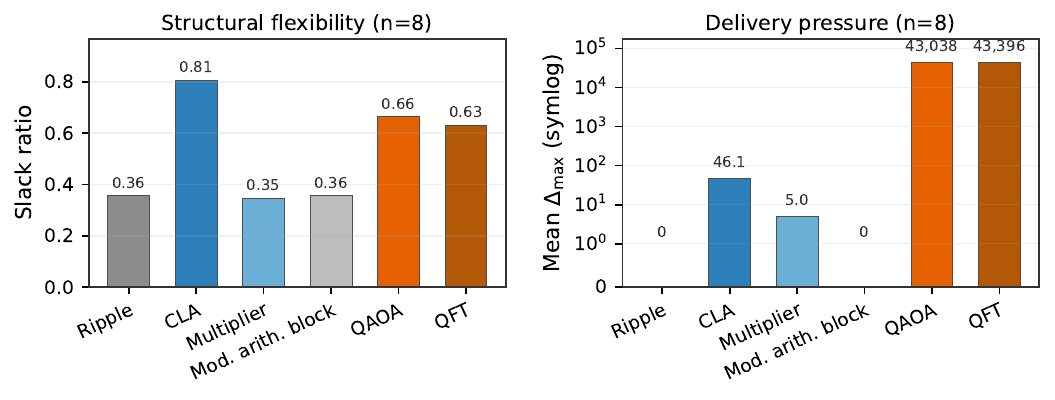}
  \caption{Representative circuit workload families in structural--system space under $\sigma_{\mathrm{static}}$ at $n=8$. The QAOA bar uses the $p=2$ dense ER MaxCut instance. The left panel reports slack ratio; the right panel reports mean $\Delta_{\max}$ in T-state units. The right panel uses a symmetric logarithmic scale to show zero-pressure, moderate-pressure, and high-pressure workloads in the same view. Ripple, CLA, multiplier, the modular-arithmetic block, and QFT use the full delivery grid; QAOA uses the reduced grid described in Section~\ref{sec:setup_workloads}.}
  \label{fig:qft_vs_real}
\end{figure}

\begin{figure}[t]
  \centering
  \includegraphics[width=\columnwidth]{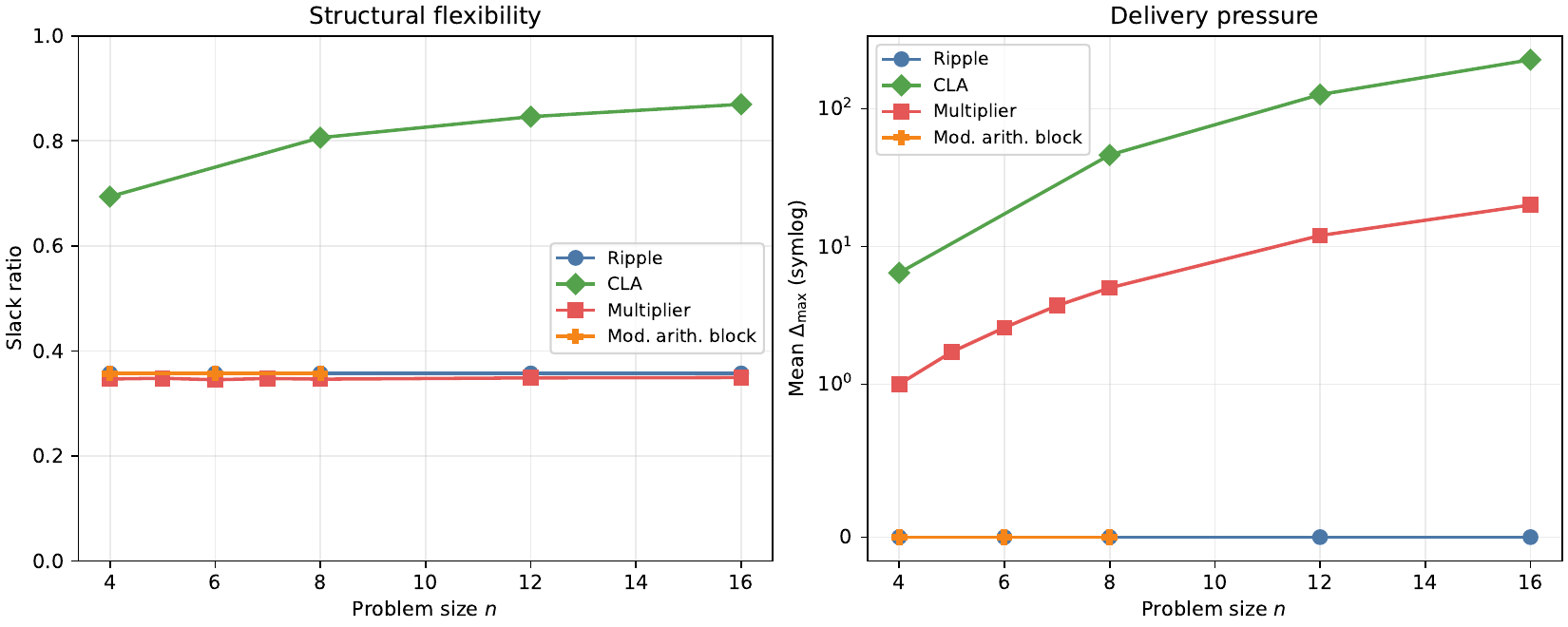}
  \caption{Scaling behavior of arithmetic and modular-arithmetic workload families under $\sigma_{\mathrm{static}}$. The left panel shows slack ratio as problem size increases; the right panel shows mean $\Delta_{\max}$ using a symmetric logarithmic scale. Ripple-carry addition and the modular-arithmetic block remain delivery-light, while carry-lookahead addition gains structural flexibility and increasing delivery pressure. QAOA is omitted from this scaling figure because its rotation-synthesis-driven delivery pressure is shown separately in Fig.~\ref{fig:qft_vs_real}.}
  \label{fig:real_scaling}
\end{figure}

Figure~\ref{fig:qft_vs_real} compares representative circuit workload families under $\sigma_{\text{static}}$: ripple-carry addition, carry-lookahead addition, multiplication, a modular-arithmetic block, QAOA MaxCut, and exact QFT. The comparison shows that delivery pressure depends strongly on circuit structure. Ripple-carry addition remains delivery-light, whereas the carry-lookahead adder evaluated here, a Kogge--Stone parallel-prefix variant~\cite{kogge1973parallel}, has much higher slack ratio and higher mean $\Delta_{\max}$ despite implementing the same arithmetic operation. The multiplier remains in a low-slack regime but exhibits nonzero delivery pressure. The modular-arithmetic block has a low-pressure profile similar to its ripple-carry building block, consistent with its building-block construction. The dense QAOA MaxCut instance also enters a high-pressure regime under the chosen Clifford$+T$ synthesis configuration. This result should be interpreted relative to the chosen Clifford$+T$ synthesis configuration: it shows that synthesis-heavy optimization workloads can create concentrated T-state demand.
Table~\ref{tab:circuit_workload_summary} summarizes the representative workload statistics underlying Fig.~\ref{fig:qft_vs_real}. High stall incidence does not always translate into large slowdown: the multiplier has frequent stalls but negligible more-than-5\% slowdown, whereas CLA and QFT show both delivery pressure and measurable slowdown. Thus, stall incidence and slowdown severity should be interpreted separately.
\begin{table}[t]
\caption{Representative circuit workload statistics under $\sigma_{\mathrm{static}}$. QAOA corresponds to the representative dense ER MaxCut instance with $p=2$ used in Fig.~\ref{fig:qft_vs_real}. Fractions are computed over each workload's evaluated $(C,B)$ grid; QAOA uses the reduced grid described in Section~\ref{sec:setup_workloads}.}
\label{tab:circuit_workload_summary}
\centering
\scriptsize
\setlength{\tabcolsep}{3pt}
\begin{tabular}{lccccc}
\hline
Workload & $n$ & Slack & Mean $\Delta_{\max}$ & Stall frac. & $>5\%$ slow. \\
\hline
Ripple & 8 & 0.357 & 0.0 & 0.009 & 0.000 \\
CLA & 8 & 0.806 & 46.1 & 0.964 & 0.063 \\
Multiplier & 8 & 0.346 & 5.0 & 0.768 & 0.000 \\
Mod. block & 8 & 0.357 & 0.0 & 0.009 & 0.000 \\
QAOA & 8 & 0.663 & $4.30{\times}10^4$ & 0.640 & 0.280 \\
QFT & 8 & 0.632 & $4.34{\times}10^4$ & 0.464 & 0.205 \\
\hline
\end{tabular}
\end{table}
Figure~\ref{fig:real_scaling} shows scaling behavior for the arithmetic and modular-arithmetic workload families. Ripple-carry addition remains stable across the evaluated range: slack ratio stays near~0.357 and mean $\Delta_{\max}$ remains near zero. The multiplier also stays in a low-slack regime, but its delivery pressure grows gradually with $n$. The carry-lookahead adder behaves differently: its slack ratio and mean $\Delta_{\max}$ both increase with problem size, showing that parallel-prefix arithmetic creates more scheduling freedom but also stronger delivery pressure under a depth-oriented schedule. The modular-arithmetic block remains close to the ripple-carry profile because its construction is based on ripple-carry building blocks.
Exact QFT remains a high-pressure transform workload. At $n=8$, QFT shows more than 5\% executable slowdown ($T_{\text{exe}} / T_{\text{static}} > 1.05$) on approximately 20.5\% of the scanned $(C,B)$ settings, but no inversion is observed in that grid. At $n=12$ and $n=16$, exact-QFT bounded-delivery scanning is omitted and those instances are included structurally only.

\begin{figure}[t]
  \centering
  \includegraphics[width=\columnwidth]{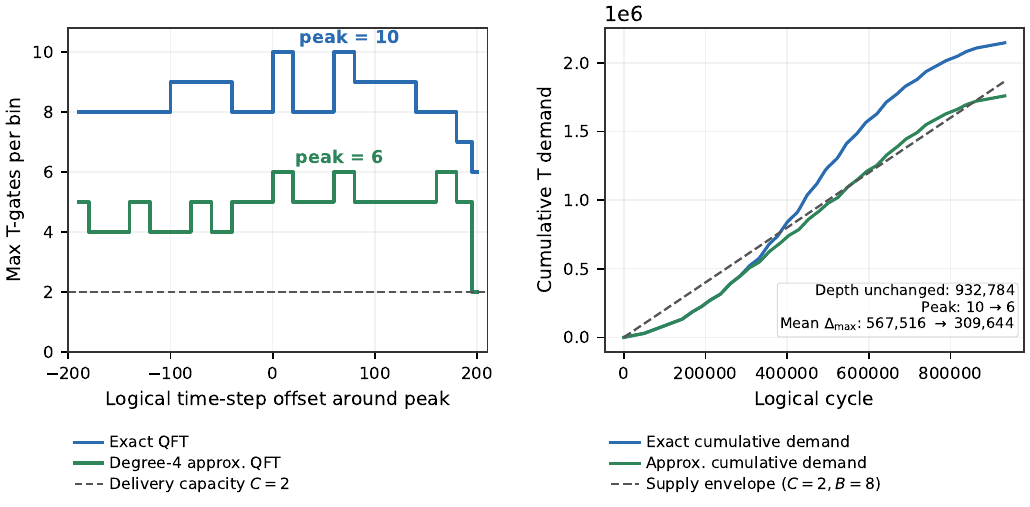}
    \caption{Exact versus degree-4 approximate QFT at $n=12$ under bounded T-state delivery. Left: binned peak-window T demand, where each bin reports the maximum number of T gates in that local interval; the dashed horizontal line indicates a representative delivery capacity ($C=2$). Approximation reduces the peak demand from 10 to 6 T gates. Right: cumulative T demand over the full logical execution compared with the supply envelope for $C=2$ and $B=8$. Although the ideal DAG depth is unchanged, approximation reduces peak demand and mean $\Delta_{\max}$, thereby lowering delivery pressure.}
  \label{fig:qft_approx}
\end{figure}

Figure~\ref{fig:qft_approx} visualizes the exact-versus-approximate QFT comparison at $n=12$. The left panel shows binned T-demand behavior around the peak-demand region, while the right panel compares cumulative demand against a representative supply envelope with $C=2$ and $B=8$. Aggregate delivery-pressure statistics are computed on the reduced grid with $C \in \{1,2,3\}$ and $B \in \{0,4,8,12\}$.

Approximation does not reduce ideal dependency depth: both circuits have DAG depth~932{,}784. It reduces T-count from 2{,}146{,}209 to 1{,}759{,}401, peak demand from 10 to 6, mean $\Delta_{\max}$ from 567{,}516 to 309{,}644, and mean slowdown ratio from 1.444 to 1.296. The fraction of settings with stalls or more-than-5\% slowdown remains unchanged on the reduced grid, so approximation primarily reduces slowdown severity rather than eliminating stalls. Circuit approximation can therefore improve executable behavior under bounded delivery even when circuit depth does not decrease~\cite{coppersmith2002approximate,ross2014optimal}.

\section{Discussion}

\subsection{Implications for Compiler Design and Architecture}
\label{sec:imp_compiler}
A first implication of these results is that T-depth minimization alone is not a sufficient compilation objective under bounded T-state delivery. The T-depth inversion cases in Section~\ref{sec:results_mismatch} show that a schedule with better static depth can still execute more slowly when T-gate demand is concentrated beyond delivery capacity. Existing transpilation heuristics in Qiskit and tket primarily optimize hardware-aware objectives such as depth, routing cost, and topology constraints. Under bounded magic-state delivery, this leaves execution time exposed to demand concentration effects that static depth does not capture. Compiler backends targeting fault-tolerant execution should therefore consider delivery capacity as an explicit scheduling constraint instead of using T-depth alone as a proxy for runtime.

A second implication is that slack ratio and $\Delta_{\max}$ support different stages of a delivery-aware compilation workflow. Slack ratio can be computed from the circuit DAG before detailed scheduling to identify circuits with exploitable scheduling freedom, while $\Delta_{\max}$ can be computed from candidate T-demand traces under the target $(C,B)$ setting to compare schedule-level delivery pressure. Together with the lower bound in Eq.~\eqref{eq:bound}, these quantities provide a lightweight way to flag schedules whose apparent T-depth advantage is likely to be lost under bounded delivery. The role of $\Delta_{\max}$ is not to replace simple capacity-aware heuristics such as $\sigma_{\text{ca}}$, but to explain when depth-oriented schedules fail and to provide a pressure metric for comparing alternatives.

A preliminary implication is that slack ratio and downstream urgency may also help order ready operations within a capacity-respecting schedule. As an additional probe reported in Appendix~\ref{app:compiler_probe}, a quota-respecting heuristic ($\sigma_{\text{quota}}$) orders ready T gates by available slack and downstream urgency. It reduces multiplier schedule length by 34--77 steps over $\sigma_{\text{ca}}$ across $n \in \{4,5,6\}$, with a positive rate of 1.00 across all evaluated settings. The adder shows no gain, suggesting that its available slack is not exploitable by this quota-respecting ordering rule. Because both policies satisfy the per-cycle T quota, $\Delta_{\max}$ and stall remain zero throughout; the observed improvement comes from critical-path ordering. This probe is not intended as a full scheduler evaluation.

Finally, bounded-delivery stalls matter not only for runtime but also for reliability-related cost. When T-state supply is delayed, logical qubits must remain active under error correction for more cycles. Although the present work does not model logical failure probabilities explicitly, each additional stall cycle extends protected idling and increases the total \textbf{space-time volume}~\cite{fowler2012surface} exposed to failure. Reducing delivery-induced backlog is therefore relevant to reliability-related cost, not just runtime.

\subsection{Limitations and Threats to Validity}
A first threat to validity is the deterministic delivery model. The execution model assumes deterministic T-state delivery at a fixed rate $C$ with a fixed buffer $B$. Real fault-tolerant systems may involve stochastic factory outputs, storage decay, and other supply variability not captured here~\cite{o2017quantum,beverland2022assessing}. The fixed-schedule lower bound is therefore established only for this deterministic model.

A second threat concerns the temporal abstraction of delivery. The main model does not explicitly include placement, routing, communication congestion, or other layout-dependent costs. The results should therefore be interpreted as isolating temporal demand--supply mismatch rather than providing a full physical-delivery model.

A third threat concerns workload coverage and external validity. The compressibility-family evaluation supports controlled predictor analysis and lower-bound validation, while the circuit workload families provide broader evidence across concrete arithmetic, optimization, and transform structures. This workload set is still not an exhaustive benchmark suite. Larger exact-QFT instances are included structurally only because Clifford$+T$ synthesis and trace-level delivery scans become expensive at high T-counts. Broader validation on larger verified arithmetic, modular-arithmetic, optimization, and simulation workloads is left for future work.

These threats define the scope in which the present results should be interpreted. Appendix~\ref{app:sensitivity} provides preliminary sensitivity checks for stochastic supply and route-induced effective-capacity effects.

\section{Conclusion and Future Work}
Static T-depth does not reliably predict executable makespan under bounded T-state delivery because execution degradation is governed by temporal demand--supply mismatch. We introduced two complementary quantities: slack ratio, a structural indicator of scheduling flexibility computed from the circuit DAG, and $\Delta_{\max}$, a schedule-level measure of delivery pressure. In the compressibility-family evaluation, $\Delta_{\max}$ is the strongest schedule-level indicator of stall and slowdown, and together with buffer capacity it induces a provable lower bound on executable makespan with zero observed violations across 4{,}904 finite schedule instances. The bound is tight in aggregate, but less so when backlog persists across multiple logical steps. Circuit workload families further show that serial arithmetic, parallel arithmetic, QAOA, and QFT occupy distinct structural--delivery regimes. The scheduling probe in Appendix~\ref{app:compiler_probe} suggests a role for slack ratio and urgency in ordering beyond delivery-pressure diagnosis. Overall, backlog-induced delay and T-depth inversion deserve explicit attention in capacity-constrained FTQC compiler design. Future work should extend the framework toward full stochastic and spatially aware delivery models, evaluate additional higher-slack FTQC workloads, and develop tighter makespan bounds for persistent-backlog cases.

\appendices

\section{Additional Analyses and Robustness Results}
\label{app:robustness}

This appendix supplements the main text with subgroup predictor stability results, lightweight sensitivity analyses, and representative positive-gap cases for the executable-makespan lower bound.

\subsection{Predictor Stability Across Subgroups}
\label{app:stability}

Table~\ref{tab:stability} reports pooled within-subgroup predictive performance for T-depth, slack ratio, and $\Delta_{\max}$. The table should be read row-wise, comparing predictors within each subgroup rather than against the slice-averaged scores reported in the main text. Across family-level and supply-regime subsets, $\Delta_{\max}$ remains the strongest predictor in nearly all nontrivial slices. Among the structural indicators, slack ratio is generally at least as informative as T-depth and is modestly stronger in the subgroups where structural differences are visible.

The low-compressibility slice is included as a control; in that regime, values such as 0.50 AUC or 0.00 correlation indicate a non-informative subgroup, not missing data.

\begin{table}[t]
\scriptsize
\setlength{\tabcolsep}{3pt}
\caption{Predictor stability across subgroups in the compressibility-family evaluation. Stall is reported as AUC and slowdown as Spearman $|\rho|$; boldface marks the best predictor in each row.}
\label{tab:stability}
\centering
\begin{tabular}{llccc|ccc}
\hline
& & \multicolumn{3}{c|}{Stall} & \multicolumn{3}{c}{Slowdown} \\
Group & Subset & T-depth & Slack & $\Delta_{\max}$ & T-depth & Slack & $\Delta_{\max}$ \\
\hline
\multirow{3}{*}{Structural}
& High & 0.500 & 0.500 & \textbf{1.000} & 0.000 & 0.000 & \textbf{0.993} \\
& Medium & 0.502 & 0.568 & \textbf{0.992} & 0.037 & 0.108 & \textbf{0.998} \\
& Low & \textbf{0.500} & \textbf{0.500} & \textbf{0.500} & \textbf{0.000} & \textbf{0.000} & \textbf{0.000} \\
\hline
\multirow{3}{*}{Capacity}
& Low C & \textbf{1.000} & \textbf{1.000} & \textbf{1.000} & 0.970 & 0.968 & \textbf{0.999} \\
& Mid C & 0.971 & 0.988 & \textbf{0.995} & 0.820 & 0.836 & \textbf{0.996} \\
& High C & 0.845 & 0.845 & \textbf{1.000} & 0.217 & 0.217 & \textbf{1.000} \\
\hline
\multirow{2}{*}{Buffer}
& Low B & 0.856 & 0.864 & \textbf{0.994} & 0.531 & 0.533 & \textbf{0.999} \\
& Mid/High & 0.857 & 0.864 & \textbf{1.000} & 0.631 & 0.637 & \textbf{0.999} \\
\hline
\end{tabular}
\end{table}

\subsection{Lower-Bound Tightness by Subset}
\label{app:bound_tightness}

Table~\ref{tab:bound_tightness_subset} reports lower-bound statistics both over all finite instances and over subsets defined by whether the buffer-adjusted surplus $P_B(\sigma)$ is zero. Aggregate tightness is high over the full finite set, but this is largely driven by instances with $P_B(\sigma)=0$. On the subset with $P_B(\sigma)>0$, the bound remains valid but substantially less tight, consistent with the multi-step backlog behavior discussed in the main text.

\begin{table}[t]
\caption{Lower-bound tightness statistics by subset in the compressibility-family evaluation.}
\label{tab:bound_tightness_subset}
\centering
\begin{tabular}{lcccc}
\hline
Subset & N & Mean gap & Within 1 cycle & Pearson $r$ \\
\hline
All finite & 4{,}904 & 0.68 & 88.9\% & 0.988 \\
$P_B(\sigma)=0$ & 4{,}522 & 0.25 & 95.6\% & 0.996 \\
$P_B(\sigma)>0$ & 382 & 5.81 & 9.69\% & 0.857 \\
\hline
\end{tabular}
\end{table}

\subsection{Sensitivity to Stochastic Supply and Route-Induced Capacity Proxies}
\label{app:sensitivity}

To probe robustness beyond the deterministic main model, we performed two lightweight sensitivity studies. 

First, we replaced the fixed per-cycle service rate with a Bernoulli-thinned service process of nominal capacity $C$ and acceptance probability $p_{\mathrm{acc}}$, where each T-state delivery attempt succeeds independently with probability $p_{\mathrm{acc}}$. This is motivated by distillation-unit acceptance abstractions used in fault-tolerant resource-estimation frameworks~\cite{beverland2022assessing}. We use $p_{\mathrm{acc}} \in \{0.99,\,0.995,\,0.999,\,1.0\}$, with $p_{\mathrm{acc}}=0.95$ as a more conservative stress case. Under this stochastic extension, the original deterministic $\Delta_{\max}$ weakens as a predictor of mean slowdown as $p_{\mathrm{acc}}$ decreases, whereas an expected-service variant (replacing $C$ with $C p_{\mathrm{acc}}$) remains substantially more stable.

Second, we modeled route-induced transport overhead through a first-order effective-capacity proxy. Recomputing the deficit with reduced effective capacity preserves the qualitative delivery-pressure ordering under moderate transport penalties. These experiments are robustness checks, not full stochastic or placement-aware architecture models.

\subsection{Representative Positive-Gap Cases}
\label{app:gap_cases}

\begin{figure}[t]
  \centering
  \includegraphics[width=\columnwidth]{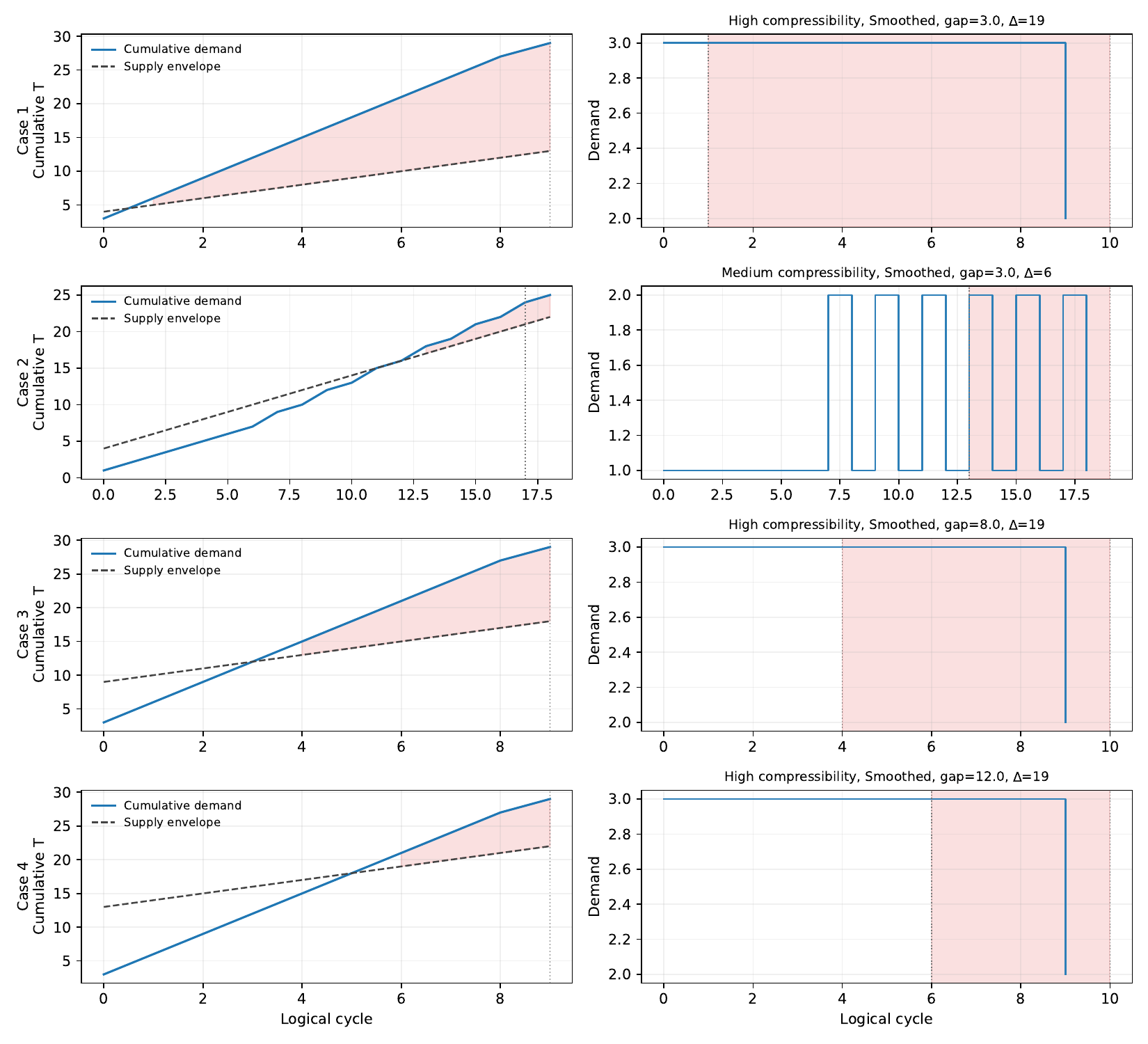}
  \caption{Representative finite positive-gap cases for the executable-makespan lower bound in the compressibility-family evaluation. In each left panel, the blue curve shows cumulative demand, the dashed line shows cumulative supply, and the red shaded region marks the backlog-active interval. In each right panel, the same red interval is projected onto the demand trace. Backlog remains active across multiple steps rather than being discharged immediately, leading to positive gaps above the lower bound.}

  \label{fig:appendix_gap_cases}
\end{figure}

Figure~\ref{fig:appendix_gap_cases} shows representative finite positive-gap cases for the executable-makespan lower bound, drawn from smoothed schedules in both high- and medium-compressibility regimes. The bound remains valid in all cases, but backlog remains active over a contiguous interval of logical steps and is not cleared immediately once the peak deficit is reached. This produces a small positive gap beyond the first-order term in Eq.~\eqref{eq:bound}, and reflects multi-step backlog activity rather than any failure of the bound itself.

\section{Probe of a Quota-Respecting Scheduling Heuristic}
\label{app:compiler_probe}

We ask whether slack ratio and urgency can improve critical-path ordering within a capacity-respecting schedule, beyond their role in delivery-pressure diagnosis. To test this, we compare $\sigma_{\text{quota}}$ against $\sigma_{\text{ca}}$ on a small set of arithmetic workloads; both policies satisfy the per-cycle T quota by construction, so any schedule-length difference reflects ordering decisions rather than reduced delivery pressure.
Table~\ref{tab:compiler_probe} reports the results.
\begin{table}[t]
\caption{Schedule-length gains of $\sigma_{\text{quota}}$ over $\sigma_{\text{ca}}$ on arithmetic workloads. Gains 
are in time steps; Pos.\ rate is the fraction of evaluated settings in which $\sigma_{\text{quota}}$ strictly improves over $\sigma_{\text{ca}}$. Static gain equals exe gain in every row because both policies satisfy the per-cycle T quota, so $\Delta_{\max}$ and stall remain zero throughout.}
\label{tab:compiler_probe}
\centering
\begin{tabular}{lccc}
\hline
Workload & $\sigma_{\text{ca}}$ len. & Schedule gain & Pos. rate \\
\hline
Adder $n=4$      & 100.0  & 0.0  & 0.00 \\
Multiplier $n=4$ & 1454.5 & 34.0 & 1.00 \\
Multiplier $n=5$ & 2298.0 & 55.0 & 1.00 \\
Multiplier $n=6$ & 3320.0 & 77.5 & 1.00 \\
\hline
\end{tabular}

\end{table}
For the adder, $\sigma_{\text{quota}}$ and $\sigma_{\text{ca}}$ produce the same schedule length, indicating that the measured slack does not create useful reordering opportunities. The multiplier gains are positive across all evaluated sizes, suggesting the advantage is systematic rather than instance-specific. These results are limited to small arithmetic circuits; broader evaluation on higher-slack workloads is left for future work.

\FloatBarrier
\section*{Acknowledgment}
ChatGPT was used for limited language editing assistance. All technical content, analysis, and final wording were reviewed and verified by the authors.

\bibliographystyle{IEEEtran}
\bibliography{ref}
\end{document}